\documentclass[english,aps,prd,floatfix,amsmath,amssymb,preprint,eqsecnum,nofootinbib,groupedaddress,titlepage]{revtex4}
\pdfoutput=1
\usepackage[T1]{fontenc}
\usepackage[latin9]{inputenc}
\usepackage[pdftex]{graphicx,color}
\usepackage{subfigure}
\usepackage{amsmath, amsfonts}
\usepackage{amssymb}
\usepackage{babel}
\usepackage{setspace} 
\begin{document}

\title{Area law violations in a supersymmetric model}

\author{Liza Huijse}
\author{Brian Swingle}
\affiliation{Department of Physics, Harvard University, Cambridge MA
02138}

\begin{abstract}
We study the structure of entanglement in a supersymmetric lattice model of fermions on certain types of decorated graphs with quenched disorder.  In particular, we construct models with controllable ground state degeneracy protected by supersymmetry and the choice of Hilbert space.  We show that in certain special limits these degenerate ground states are associated with local impurities and that there exists a basis of the ground state manifold in which every basis element satisfies a boundary law for entanglement entropy.  On the other hand, by considering incoherent mixtures or coherent superpositions of these localized ground states, we can find regions that violate the boundary law for entanglement entropy over a wide range of length scales.  More generally, we discuss various desiderata for constructing violations of the boundary law for entanglement entropy and discuss possible relations of our work to recent holographic studies.
\end{abstract}

\maketitle

\section{Introduction}

Entanglement plays a fundamental role in quantum many-body physics.
It can be used to identify topological phases of matter not characterized
by any broken symmetry \cite{topent1,topent2} as well as gapless critical points \cite{geo_ent,eeqft,vidal_qcp,ON_ent,decon_qcp_ee}. The structure
of many-body entanglement has led to a wide variety of promising
new variational states \cite{vidal_mera,peps,terg} and can be understood as underlying the functioning
of DMRG. Entanglement also provides a new perspective on global issues
about the landscape of quantum phases, including questions about the
structure of renormalization group flows \cite{holo_c_2}. Despite these many promising
achievements, we are still exploring the basic structure of entanglement
and do not yet have a general picture of many-body entanglement. Even
a basic question like the scaling of entanglement with system size
remains open despite many recent advances. In these notes we describe
a large class of models built using the technology of supersymmetry
where we can construct states with tunable entanglement properties.

The entanglement entropy, defined as the von Neumann entropy of the
reduced density matrix of a spatial subsystem, typically obeys an
area law for many-body ground states of local Hamiltonians \cite{arealaw1}. That is,
$S_{L}=-{\rm Tr}(\rho_{L}\log\rho_{L})\sim L^{d-1}$, where $L$ is the
linear size of the spatial subregion and $d$ the spatial dimension
of the system. There are, however, examples of gapless states that
violate the area law logarithmically, i.e., $S_{L}\sim L^{d-1}\log L$.
The most well-known example being the Landau Fermi liquid ground state \cite{fermion1,fermion2,fermion3,bgs_ferm1}.  There has also been considerable recent interest in violations of the area law from the point of view of holographic duality \cite{2011arXiv1111.1023O,PhysRevB.85.035121,Dong12}.
Under certain constraints, the most important being the absence of
long-range interactions, ground state degeneracy, and infinite fine-tuning,
it seems that the area law is violated at most logarithmically at the largest scales \cite{Swingle11}.

When these constraints are relaxed, however, there are constructions that lead to ground states with an anomalous violation of the area law. For instance, when long-range interactions (specifically long range hoppings) are allowed it is possible to construct one dimensional systems for which the entanglement entropy of a pure ground state scales as any function of $L$ less than extensive \cite{inf_fermipts}. Another possibility is to have a large ground state degeneracy, for instance models with an extensive ground state entropy (see e.g. \cite{Misguich03,fendley-2005-95,vanEerten05}) are likely to also have an extensive entanglement entropy. Finally, models with entanglement entropy scaling in the whole range between area law and extensive were recently constructed in the context of holography \cite{Dong12}. For the latter models it is hard to say which of the above constraints are violated. Furthermore, it is not well understood in what state the entanglement entropy is computed and if the anomalous scaling persists to all scales in the infrared. However, due to the generality of the construction, the answer to these and related questions may well provide further guidelines as to what type of entanglement properties may be expected.

The way anomalous violations of the area law may arise in systems with a large ground state degeneracy can be understood as follows.
Let $\{|\alpha\rangle\}$ be the basis spanning the ground state subspace
of the Hilbert space. It follows that a typical ground state is a
cohorent state described by the density matrix $\rho_{c}=\sum_{\alpha,\beta}c_{\alpha}c_{\beta}|\alpha\rangle\langle\beta|$.
For sufficiently orthogonal states, $|\alpha\rangle$, the reduced
density matrix, obtained by tracing out a spatial subregion, will
be very similar to the density matrix of a completely mixed state.
For systems with a large ground state degeneracy we thus conclude
that the entanglement entropy can grow as $\sim\log N_{{\rm gs}}(L)+S_{\alpha}$,
where $N_{{\rm gs}}(L)$ is the number of ground states in a system of
linear size $L$ and $S_{\alpha}$ is the entanglement entropy of
the typical ground state $|\alpha\rangle$. Typically $S_{\alpha}$
obeys the area law. For a system with a ground state degeneracy that
is exponential in the volume, the entanglement entropy of the coherent
state is thus completely dominated by the first part: $\log N_{{\rm gs}}\sim L^{d}.$
From this perspective, for a system to have an entanglement entropy
that scales $\sim L^{d-a}$, with $0<a<1$ , we need the ground state
degeneracy to be exponential in $L^{d-a}$. To our knowledge all known
examples of local Hamiltonians with a large ground state degeneracy
have either $a=0$ or $a=1$ (see e.g. \cite{Huijse08a}).

A striking example of a model with a large ground state degeneracy
is the supersymmetric model for strongly interacting itinerant fermions
\cite{fendley-2003-90} (for a review see \cite{HuijseT10}). The
Hamiltonian is supersymmetric by construction: it is defined as the
anti-commutator of two supercharges. This construction allows for
an increased analytic control of a lattice model for strongly interacting
fermions. In particular, it is often possible to compute the ground
state degeneracy exactly by solving the cohomology problem. From this
model a large variety of ground state degeneracies is observed, ranging
from a unique ground state to a ground state degeneracy exponential
in the number of sites of the system. It is thus natural to ask, as suggested in \cite{PhysRevB.85.035121}, if
there are supersymmetric lattice models with a ground state degeneracy
that allows for an anomalous violation of the area law.

Here we construct a class of graphs on which the supersymmetric lattice
model has a ground state degeneracy that can be controlled at will.
In particular, if the graph has volume $L^{d}$, with $d$ the dimension
of the graph, the ground state degeneracy grows as $\sim2^{t}$, where
$t$ can be any number between $0$ and $L^{d}$.

This work is organized as follows. First, we present a set of criteria
that, for our taste, should be obeyed when searching for violations
of the boundary law. Second, we present our supersymmetric lattice
model and analyze in detail its entanglement properties. Finally,
we discuss in the larger context the virtues and defects of our construction.

\section{Model Criteria}

It is widely believed that gapped phases in higher dimensions satisfy
a boundary law for entanglement entropy, but even this ``obvious''
statement is unproven outside the one dimensional arena. In searching
for violations of the boundary law, we want to set out certain basic
rules. These rules represent an attempt to select microscopic models
that are realistic and generic for quantum many-body physics.
\begin{itemize}
\item {{[}Short-range interactions{]}} We only consider systems without
long-range interactions. Of course, gapless modes are allowed, but
we do not ``integrate them out'' to produce long-range
interactions for the remaining degrees of freedom. If this requirement
is relaxed, it is known that one can obtain a plethora of entanglement
scalings. One cause for concern is the Coulomb interaction which,
in insulating phases, is long ranged (although this is quite different from having long range hoppings).  In appendix \ref{app:fractalFS} we give an alternative fermion construction of a state with anomalous entanglement entropy using long-range hoppings.
\item {{[}Dimensionality{]}} We only consider systems that can be ascribed
an integer spatial dimensionality. We do not consider fractal graphs
or other spaces with unusual spatial structure e.g. hyperbolic space.
It is not that these spaces are not interesting, only that the notion
of boundary law violation becomes less meaningful.
\item {{[}Randomness{]}} Our construction will make use of non-translation
invariant graphs or quenched randomness which is of a rather peculiar
type. While we would like to have models that do not rely on this
physics, realistic materials do have impurities, so although the type
of randomness we consider is unusual, quenched randomness is present
in general.
\item {{[}Fine-tuning{]}} We would like to avoid as much as possible fine-tuning
of the Hamiltonian. Any finite amount of fine-tuning can be interpreted
as some kind of multicritical phenomenon, but with an infinite amount
of fine tuning we can also easily construct models with anomalous
entanglement e.g. spin chains where every spin in a segment forms
a singlet with a spin outside the segment.
\end{itemize}
Our construction is a supersymmetric lattice model in any dimension
with short range interactions and dilute randomness. Except for a
vanishing density of impurities, it behaves like a conventional quantum
system in integer dimension. Furthermore, although the supersymmetry
implies some degree of fine tuning, it is not infinitely fine tuned.

\section{The model}

Let us now define the model. The degrees of freedom are spinless fermions
living on the square lattice. A fermion at site $i$ is created by
the operator $c_{i}^{\dagger}$ with $\{c_{i},c_{j}^{\dagger}\}=\delta_{ij}$.
The fermions have a hard core, meaning that they are not only forbidden
to be on the same site as required by Fermi statistics, but are also
forbidden to be on adjacent sites. Their creation operator is $d_{i}^{\dagger}=c_{i}^{\dagger}P_{\langle i\rangle}$,
where $P_{<i>}=\prod_{j\textrm{ next to }i}(1-c_{j}^{\dag}c_{j})$
is zero if any site next to $i$ is occupied. The Hamiltonian is defined
in terms of the supercharges, $Q=\sum_{i}d_{i}^{\dagger}$ and $Q^{\dagger}=\sum_{i}d_{i}$,
that obey $Q^{2}=(Q^{\dagger})^{2}=0$. The Hamiltonian is
\begin{equation}
H=\{Q^{\dag},Q\}=\sum_{\langle i,j\rangle}d_{i}^{\dag}d_{j}+\sum_{i}P_{\langle i\rangle}.
\end{equation}
 From its definition it follows that the Hamiltonian commutes with
the supercharges. Imposing this structure has some immediate consequences:
supersymmetric theories are characterized by a positive definite energy
spectrum and a twofold degeneracy of each non-zero energy level. The
two states with the same energy are called superpartners and are related
by the supercharge. In other words, all eigenstates with an energy
$E_{s}>0$ form doublet representations of the supersymmetry algebra.
A doublet consists of two states $(|s\rangle,Q|s\rangle)$, such that
$Q^{\dag}|s\rangle=0$. Finally, all states with zero energy must
be singlets: $Q|g\rangle=Q^{\dag}|g\rangle=0$ and conversely, all
singlets must be zero energy states \cite{Witten82}.

\subsection{Cohomology}

For the supersymmetric models, cohomology has proven to be a very
powerful tool to extract information about the zero energy ground
state(s) of the models (see for example \cite{fendley-2003-90,fendley-2005-95,Jonsson05p,Huijse08b,Huijse10,HuijseT10}). The key ingredient is the fact that ground
states are singlets; they are annihilated both by $Q$ and $Q^{\dag}$.
This means that a ground state $|g\rangle$ is in the kernel of $Q$:
$Q|g\rangle=0$ and not in the image of $Q$, because if we could
write $|g\rangle=Q|f\rangle$, then $(|f\rangle,|g\rangle)$, would
be a doublet. So the ground states span a subspace $\mathcal{H}_{Q}$
of the Hilbert space $\mathbf{H}$ of states, such that $\mathcal{H}_{Q}=\ker Q/\textrm{Im}Q$.
This is precisely the definition of the cohomology of $Q$. So the
ground states of a supersymmetric theory are in one-to-one correspondence
with the cohomology of $Q$. It follows that the solution of the cohomology
problem gives the number of zero energy states for each particle number
sector. Equivalently, we find that zero energy states are in one-to-one
correspondence with the homology elements of $Q^{\dagger}$.

We compute the cohomology using the `tic-tac-toe' lemma of \cite{BottTu82}.
This says that under certain conditions, the cohomology $\mathcal{H}_{Q}$
for $Q=Q_{1}+Q_{2}$ is the same as the cohomology of $Q_{1}$ acting
on the cohomology of $Q_{2}$. In an equation, $\mathcal{H}_{Q}=\mathcal{H}_{Q_{1}}(\mathcal{H}_{Q_{2}})\equiv\mathcal{H}_{12}$,
where $Q_{1}$ and $Q_{2}$ act on different sublattices $S_{1}$
and $S_{2}$. We find $\mathcal{H}_{12}$ by first fixing the configuration
on all sites of the sublattice $S_{1}$, and computing the cohomology
$\mathcal{H}_{Q_{2}}$. Then one computes the cohomology of $Q_{1}$,
acting not on the full space of states, but only on the classes in
$\mathcal{H}_{Q_{2}}$. A sufficient condition for the lemma to hold
is that all non-trivial elements of $\mathcal{H}_{12}$ have the same
$f_{2}$ (the fermion-number on $S_{2}$).

Although ground states are in one-to-one correspondence with cohomology
elements, the two are not equal unless the cohomology element happens
to be a harmonic representative of the cohomology. Harmonic representatives
are elements of both the cohomology of $Q$ and the homology of $Q^{\dag}$.
So they are annihilated by both supercharges, which is precisely the
property of a zero energy state. It follows that, although the solution
of the cohomology problem gives the number of ground states, it typically
does not give the ground states themselves. For a more leisurely introduction
to cohomology and an exposition of the relation between the supersymetric
model and independence complexes we refer the reader to \cite{Huijse10,HuijseT10}.

We now briefly state the cohomology results for the one dimensional
chain \cite{fendley-2003-90}, because we will use these results many
times throughout the paper. For the periodic chain of length $L=3n+a$
the cohomology is trivial for all particle numbers, $f$, except for
$f=n$, where we have
\begin{equation}
{\rm dim}(\mathcal{H}_{Q})=\begin{cases}
1 & \textrm{for \ensuremath{a=\pm1},}\\
2 & \textrm{for \ensuremath{a=0}.}
\end{cases}
\end{equation}
 Similarly, for open boundary conditions we have
\begin{equation}
{\rm dim}(\mathcal{H}_{Q})=\begin{cases}
1 & \textrm{for \ensuremath{a=0} and \ensuremath{a=-1},}\\
0 & \textrm{for \ensuremath{a=1},}
\end{cases}
\end{equation}
 for a chain of length $L=3n+a$ and $n$ particles and ${\rm dim}(\mathcal{H}_{Q})=0$
at all other particle numbers. In particular, we note that the cohomology
of an isolated site that can be both empty and occupied ($L=1$ with
open boundary conditions) is trivial. This is equivalent to the statement
that the single site chain has no zero energy states, indeed the empty
and occupied state form a doublet of energy $E=1$.

\subsection{Decorated graphs with quenched disorder}

\begin{figure}
\centering
  \subfigure[The original graph $\Lambda$]{\includegraphics[width=0.3\textwidth]{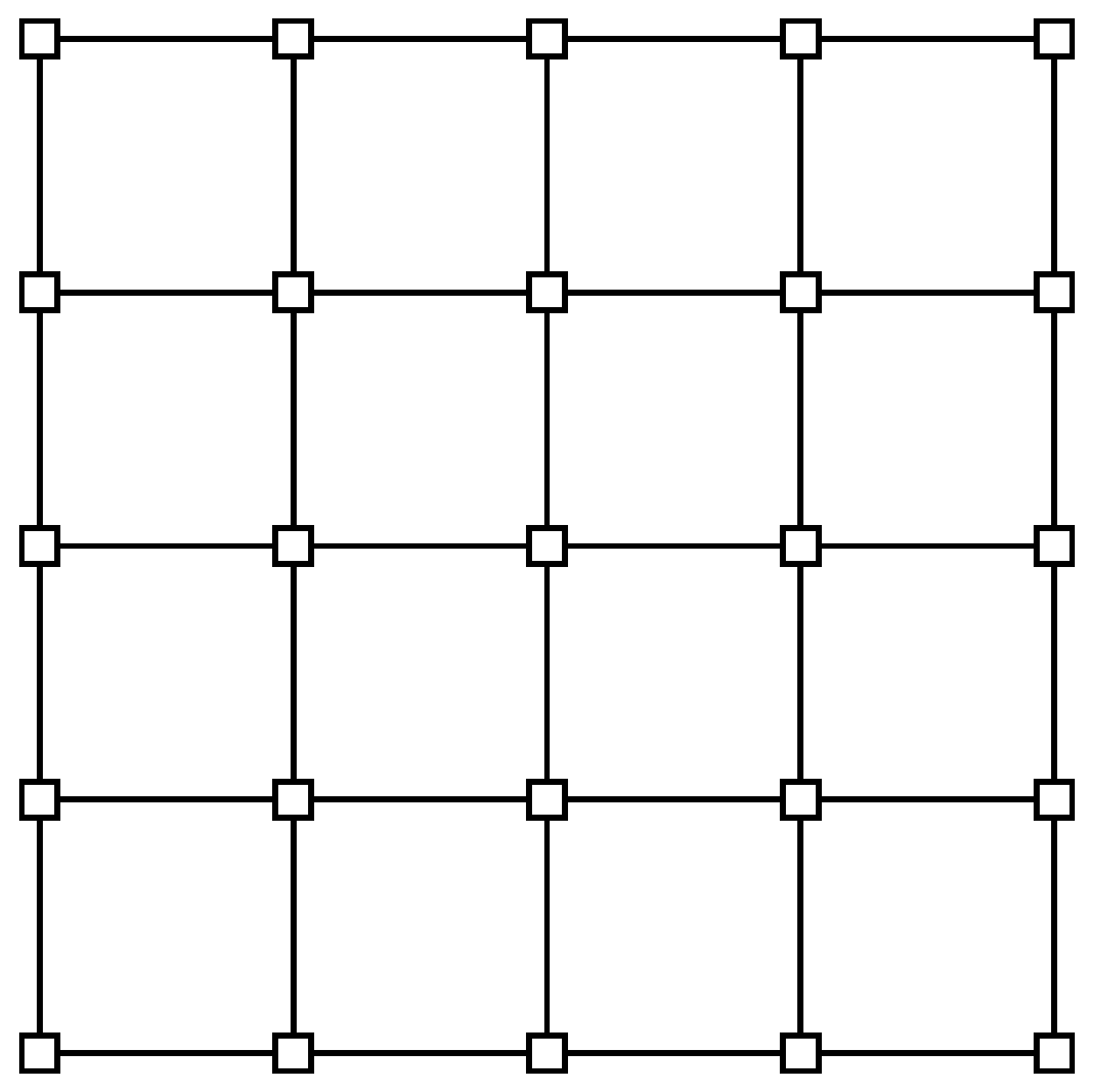}}
    \hspace{0.3cm}
  \subfigure[The decorated graph $\Lambda_3$]{\includegraphics[width=0.3\textwidth]{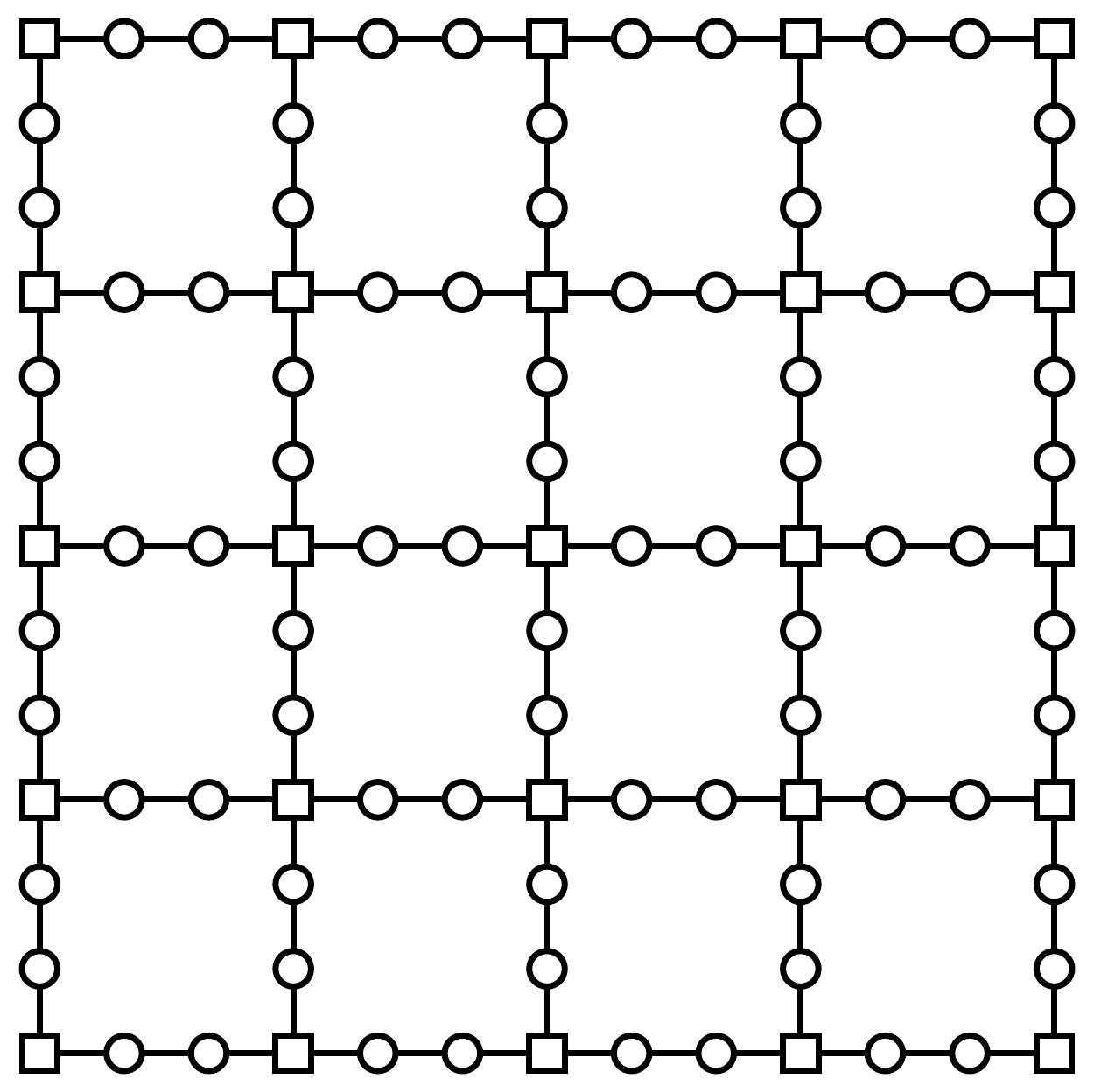}}
    \hspace{0.3cm}
  \subfigure[The decorated graph with quenched disorder $\Lambda_{3,t}$ with $t=6$]{\includegraphics[width=0.3\textwidth]{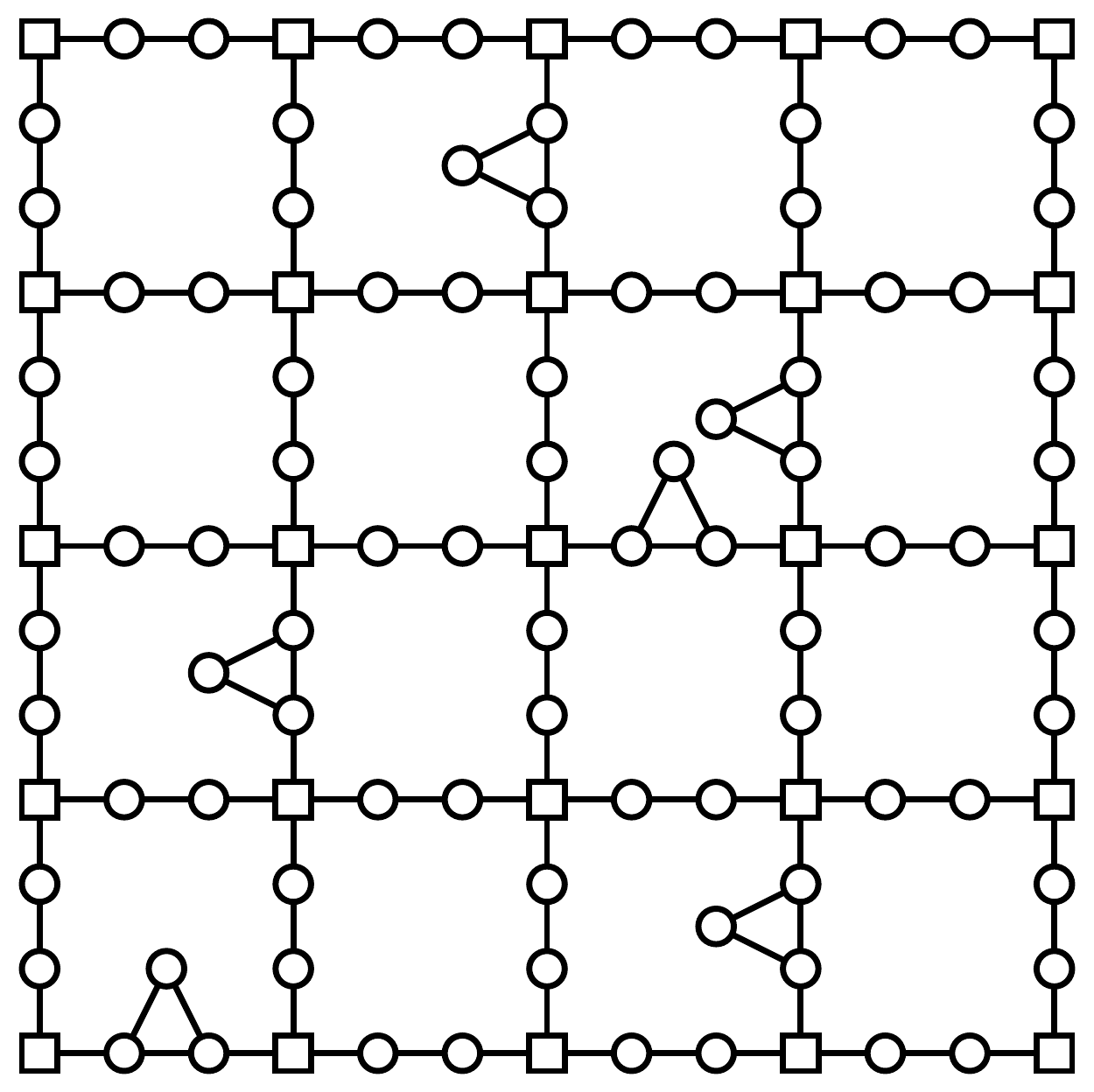}}
  \caption{A decorated graph with quenched disorder is constructed from the square lattice via the procedure explain in the text. The sites of subgraphs $S_1$ and $S_2$ are depicted as squares and circles, respectively.}
  \label{fig:graphs}
\end{figure}

The graphs we consider in this work are constructed as follows. We
start with an original graph, $\Lambda$, with $N_{\Lambda}$ vertices
and $L_{\Lambda}$ links. The only restriction on the graph, $\Lambda$,
is that it does not contain disconnected subgraphs. Now construct
the graph, $\Lambda_{3}$, from the original graph by adding 2 additional
vertices on every link. Then from $\Lambda_{3}$ the decorated graph,
$\Lambda_{3,t}$, is constructed by adding $t$ additional sites to
the graph and connecting each added site to a unique pair of sites
on a link of the original graph, $\Lambda$. If we call the collection
of sites on the original graph, $\Lambda$, the subgraph $S_{1}$
and all the added sites subgraph $S_{2}$, we find that $S_{2}$ is
a collection of $L_{\Lambda}-t$ open 2-site chains and $t$ periodic
3-site chains. An example is shown in figure \ref{fig:graphs}.

\section{Ground state degeneracy}

In this section we compute the ground state degeneracy of the supersymmetric
model on the decorated graphs with quenched disorder, $\Lambda_{3,t}$.
The supersymmetric model on the decorated graphs without disorder,
$\Lambda_{3}$, is discussed in detail in \cite{Huijse11b}. The cohomology
problem is easily solved and the ground state is found to be twofold
degenerate \cite{fendley-2005-95}.

As a warm-up we briefly review this computation. Consider the original
graph, $\Lambda$, as the subgraph $S_{1}$ and the additional sites
as the subgraph $S_{2}$. The subgraph $S_{2}$ is a collection of
two site chains. Remember that an isolated site that can be both empty
and occupied has a trivial cohomology. It follows that the non-trivial
elements in $\mathcal{H}_{Q_{2}}$ have the $S_{1}$ sites neighboring
a two site chain on $S_{2}$ either both empty or both occupied. Consequently,
there are only two non-trivial elements in $\mathcal{H}_{Q_{2}}$,
one with $S_{1}$ completely empty and one with $S_{1}$ completely
filled. If we leave $S_{1}$ completely empty, we obtain one non-trivial
element of $\mathcal{H}_{Q_{2}}$ in the sector with $L_{\Lambda}$
fermions, where $L_{\Lambda}$ denotes the number of links in the
original graph $\Lambda$. The element with $S_{1}$ completely filled,
clearly has fermion number $N_{\Lambda}$, where $N_{\Lambda}$ denotes
the number of vertices in the original graph $\Lambda$. It now quickly
follows that these two elements are also in $\mathcal{H}_{12}$, since
within $\mathcal{H}_{Q_{2}}$ both states are in the kernel of $Q_{1}$
and not in the image of $Q_{1}$. Now remember that for the 'tic-tac-toe'
lemma to hold a sufficient condition was that all elements in $\mathcal{H}_{12}$
have the same number of fermions on sublattice $S_{2}$. Here this
condition is clearly not met. One can easily show, however, that the
lemma also holds when the elements of $\mathcal{H}_{12}$ do not differ
in their total fermion number by one. We thus conclude that $\mathcal{H}_{12}=\mathcal{H}_{Q}$
provided that $N_{\Lambda}\neq L_{\Lambda}\pm1$.

As an example consider the square lattice with doubly periodic boundary
conditions as the original graph $\Lambda$. We find that $L_{\Lambda}=2N_{\Lambda}$
and the total number of sites in $\Lambda_{3}$ is $N=2L_{\Lambda}+N_{\Lambda}$.
Consequently, this lattice has one ground state at $1/5$ filling
and one at $2/5$ filling.

Let us now consider the cohomology problem on the graphs $\Lambda_{3,t}$.
As before we define the subgraph $S_{1}$ as the original graph $\Lambda$
and the subgraph $S_{2}$ as the rest of the sites. It follows that
$S_{2}$ is a collection of $L_{\Lambda}-t$ open 2-site chains and
$t$ periodic 3-site chains. Note that a periodic 3-site chain has
two zero energy ground states with fermion number one. It follows
that the configuration with all $S_{1}$ sites empty corresponds to
$2^{t}$ non-trivial elements in $\mathcal{H}_{Q_{2}}$. Furthermore,
we note that if two $S_{1}$ sites neighboring a 3-site chain are
occupied, the 3-site chain effectively becomes an isolated site. Therefore
the configuration with all $S_{1}$ sites occupied is not an element
of $\mathcal{H}_{Q_{2}}$. However, if only one of the two $S_{1}$
sites neighboring a 3-site chain is occupied, the 3-site chain effectively
becomes a 2-site chain, which does have a zero energy state. It follows
that other configurations on $S_{1}$ may correspond to non-trivial
elements in $\mathcal{H}_{Q_{2}}$. This complicates the computation
of $\mathcal{H}_{12}$, however, it is easy to verify that the configuration
with all $S_{1}$ sites empty is the only allowed configuration in
$\mathcal{H}_{Q_{2}}$ under the following condition. The 3-site chains
are distributed over the graph in such a way that any two $S_{1}$
sites neighboring a 3-site chain can be connected to each other by
drawing a path that contains only 2-site chains on $S_{2}$ and $S_{1}$
sites. Although it may be possible to solve the cohomology problem
in other cases as well, we do not consider these here, since the above
constraint is acceptable for the purposes of the present work. Under
the above condition, it immediately follows that all $2^{t}$ non-trivial
elements of $\mathcal{H}_{Q_{2}}$ are also elements of $\mathcal{H}_{Q}$,
since they are all in the kernel of $Q_{1}$ within $\mathcal{H}_{Q_{2}}$
and they all have the same number of fermions on $S_{2}$.

We conclude that the ground state degeneracy of the supersymmetric
model on the graph $\Lambda_{3,t}$ is $2^{t}$ ($t>0$) under the
following conditions:
\begin{itemize}
\item The original graph, $\Lambda$, does not contain disconnected subgraphs,
\item The original graph, $\Lambda$, has $N_{\Lambda}$ vertices and $L_{\Lambda}$
links such that $N_{\Lambda}\neq L_{\Lambda}\pm1$,
\item The $t$ periodic 3-site chains are distributed over the graph in
such a way that any two $S_{1}$ sites neighboring a 3-site chain
can be connected to each other by drawing a path that contains only
2-site chains on $S_{2}$ and $S_{1}$ sites.
\end{itemize}
It is easily verified that these conditions can be met for $t$ in
the entire range from order one to order $L^{d}$.

\section{Ground states for large and small staggering}\label{sec:stag}

For the supersymmetric model on $\Lambda_{3}$ graphs in the presence
of a staggering parameter, $y$, it was found that in the limit of
small and large staggering the ground states take on a simple form
\cite{Huijse11b}. In particular, the ground state with $L_{\Lambda}$
fermions is a valence bond solid state in the limit of large staggering
and a quantum liquid state in the limit of small staggering. The quantum
liquid state can be expressed as a projected valence bond solid state.
In this section we will briefly review there results and show that
in the presence of quenched disorder similar expressions for the $2^{t}$
ground states can be found in the small and large staggering limits.

We introduce the staggering parameter, $y$, by redefining the supercharge
as

\begin{equation}
Q=Q_{1}+yQ_{2},
\end{equation}
and similar for $Q^{\dagger}$, where $Q_{i}$ acts solely on the
subgraph $S_{i}$. The Hamiltonian then reads $H=\{Q,Q^{\dag}\}=H_{0}+yH_{1}+y^{2}H_{2}$,
where $H_{0}$ is purely diagonal, $H_{1}$ is purely off-diagonal,
it is the hopping between staggered and non-staggered sites, and finally,
$H_{2}$ consist of a diagonal and an off-diagonal part, where the
off-diagonal part represents hopping between the added $S_{2}$ sites
on a link of the graph $\Lambda$.

The limit of $y\to\infty$ is easy. To zeroth order in $1/y$ the
Hamiltonian reads $H/y^{2}\approx H_{2}=\{Q_{2}^{\dag},Q_{2}\}$.
The two elements of $\mathcal{H}_{Q_{2}}$, one with $S_{1}$ completely
empty and one with $S_{1}$ completely filled, are precisely the ground
states of this Hamiltonian. We focus on the state with $L_{\Lambda}$
particles, that is the state with $S_{1}$ completely empty. The $S_{2}$
subgraph is a collection of two-site chains. The ground state of a
two-site chain contains one particle that resonates between the two
sites: $|\psi_{0}\rangle=(c_{1}^{\dag}-c_{2}^{\dag})/\sqrt{2}|\emptyset\rangle$.
It follows that in the state with $S_{1}$ completely empty there
is one particle on each two-site chain of $S_{2}$ sites that resonates
between the two sites. We denote this product state by $|{\rm VBS}\rangle$
inspired by the obvious similarity to a valence bond solid state.

It is clear that this result carries over directly to the decorated
graphs with quenched disorder. On the triangles there are now two
single particle states that are annihilated by $H_{2}$. In the limit
of $y\to\infty$ there is a basis of the $2^{t}$ dimensional ground
state subspace in terms of valence bond solid states characterized
by the internal states of the triangles. Let this basis be $|\{\alpha_{\ell}\}\rangle$
where $\alpha_{\ell}$ labels the state of the triangle on link $\ell$.

The limit of small staggering is more involved. For $y=0$ we have
$Q=Q_{1}$ and we should solve the cohomology problem by looking at
$\mathcal{H}_{Q_{1}}$ directly. Since $S_{1}$ is a collection of
disconnected sites, the non-trivial cohomology elements of $\mathcal{H}_{Q_{1}}$
are all configurations which have at least one neighboring site of
each $S_{1}$ site occupied. There are many such configurations, with
particle numbers in the whole range between $N_{\Lambda}$ and $L_{\Lambda}$.
So for $y$ strictly zero, we thus find that there is a large ground
state degeneracy. It can be shown, however, that this degeneracy is
lifted at second order in degenerate perturbation theory \cite{Huijse11b}.
The effective Hamiltonian at order $y^{2}$ that couples the zeroth-order
ground states in the sector with $L_{\Lambda}$ particles can be written
as

\begin{equation}
H_{{\text{eff}}}=\sum_{{\text{links}}}\sum_{{\tau,\tau'}}\tilde{P}|\tau\rangle\langle\tau'|\tilde{P}-\sum_{{\text{vertices}}}|1\rangle\langle1|,
\end{equation}
where the sums are over the links and vertices of the original graph
$\Lambda$, $\tau$ runs over the possible single particle configurations
on the $S_{2}$ sites added to a single link of the original graph
$\Lambda$. By the state $|1\rangle$ we denote the vertex configurations
in which precisely one neighboring site is occupied. Finally, $\tilde{P}$
projects onto the zeroth-order ground state subspace, that is, it
projects out all configurations that contain vertices with all neighboring
sites empty. For the decorated graph, $\Lambda_{3}$, this effective
Hamiltonian has precisely one zero energy ground state given by $\tilde{P}|{\rm VBS}\rangle$%
\footnote{In \cite{Huijse11b} this state is identified as a quantum liquid
state. This can be seen by absorbing the signs of the coefficients
in the wavefunction into the definition of the fermion creation and
annihilation operators. It then follows that the projected valence
bond solid state is an equal amplitude superposition of all zeroth-order
ground states.%
}.

It is instructive for the following to see why the projected VBS state
is annihilated by the effective Hamiltonian. First consider the action
of the first term and focus on a single link. If both states $\tau$
on that link are allowed then the singlet state is annihilated by
$\sum_{\tau,\tau'}|\tau\rangle\langle\tau'|$. However, if only one
state is allowed, then the action of $\sum_{\tau,\tau'}\tilde{P}|\tau\rangle\langle\tau'|\tilde{P}$
is purely diagonal. It is not hard to see that this happens when the
state $\tau$ is such that the particle is the only particle neighboring
the adjacent $S_{1}$ site, that is the neighboring vertex is in state
$|1\rangle$. It follows that the action of the first term in the
effective Hamiltonian on the projected VBS state can be written as
$\sum_{{\text{vertices}}}|1\rangle\langle1|$, which precisely cancels
the second term.

For the decorated graphs with quenched disorder in the limit of small
staggering the above arguments basically go through unaltered, with
the exception that the effective Hamiltonian now has $2^{t}$ zero
energy states that can be written in a maximally localized form as
$\tilde{P}|\alpha\rangle$, where $|\alpha\rangle \in |\{\alpha_{\ell}\}\rangle$
is a VBS state characterized by the states of the triangles. Indeed, one readily
verifies that the effective Hamiltonian annihilates all these projected
VBS states, the only additional check we need to do is to make sure
that these states span the full $2^{t}$ dimensional subspace. We
check this by verifying that $\tilde{P}$ has full rank on the subspace
spanned by the VBS basis $|\{\alpha_{\ell}\}\rangle$. It is clear
that a linear superposition of the VBS states is annihilated by $\tilde{P}$
if and only if it consists exclusively of configurations that contain
vertices with all neighboring sites empty. Since the VBS states differ
only in the states on the triangles, we can choose a superposition
of VBS states such that on some link $\ell$ the particle sits on
the $S_{2}$ site that has no neighboring $S_{1}$ sites. We could
therefore obtain a superposition of VBS states that consists exclusively
of configurations where one vertex has all neighboring sites empty,
if all links coming out of that vertex contain triangles. This last
condition, however, is in contradiction with the condition we imposed
on the distribution of triangles over the graph $\Lambda$. We thus
conclude that $\tilde{P}$ has full rank on the subspace spanned by
the VBS states. It follows that the projected VBS states span the
full $2^{t}$ dimensional ground state subspace.

We conclude that in the limit of small and large staggering there is a maximally localized basis given by $\tilde{P}|\{\alpha_{\ell}\}\rangle$ and $|\{\alpha_{\ell}\}\rangle$, respectively, that spans the $2^{t}$ dimensional ground state subspace of the decorated graphs with quenched disorder, $\Lambda_{3,t}$. For intermediate values of the staggering parameter a simple physical picture of the ground states is lacking.

Finally, before we go on the consider the entanglement entropy of a typical ground state of the model with quenched disorder, we briefly discuss the entanglement properties of the (projected) VBS states.  It is clear that each unprojected VBS state satisfies a boundary law for entanglement entropy.  The entropy is simply given by the number of valence bonds cut by the entangling surface times the elementary entropy of a single valence bond.  The only issue is the projection.  We have already shown that the projector $\tilde{P}$ has no kernel acting on the VBS states.  The other important ingredient is to determine the overlap between two different projected VBS states.  Given the gapped and short ranged nature of the VBS states, this overlap should be exponentially decaying in distance.  In other words, two projected VBS states distinguished by configurations of spatially separated triangles will be orthogonal up to corrections exponential in the distance between the triangles.  With this additional fact, we claim the projected VBS states also satisfy a boundary law.  This is because after we orthonormalize the projected states we must only mix in exponentially small amounts of states associated with distant triangles, so we do not expect any qualitative change in the degree of entanglement within a given localized state.

\section{Synthesis}

We have shown that the supersymmetric model on the graph $\Lambda_{3,t}$
has $2^{t}$ exact ground states.  Once we include the deformation parameter $y$, we can provide model wavefunctions for these states at small and large $y$. For large $y$ the ground states are simply VBS states, while for small $y$ the wavefunctions are projected versions of the same VBS states. In either limit there
is a basis of the $2^{t}$ dimensional ground state subspace in which
the basis vectors are maximally localized e.g. associated with local
changes in the states of triangles. Let this basis be $|\{\alpha_{\ell}\}\rangle$
where $\alpha_{\ell}$ labels the state of the triangle on link $\ell$.
As demonstrated above, each of these states satisfies a boundary law
for entanglement entropy. Except for the complication of the projection,
entanglement is due only to the short range valence bonds cut by the
entangling surface. Let us now ask what sort of combinations of these
local states should be allowed. Let $A$ be a spatial region in the
graph not necessarily consisting of sites from only $S_{1}$ or $S_{2}$
and let $B$ be its complement. From the perspective of cooling to
the ground state, a natural choice is the incoherent state
\begin{equation}
\rho_{AB}=\frac{1}{2^{t}}\sum_{\alpha}|\alpha\rangle\langle\alpha|.
\end{equation}
 This state is already mixed at the level of the total system and
hence much of the entropy of $\rho_{A}$ is due to classical effects
and not entanglement. Nevertheless, the entropy of $A$ in this state
goes like $t\log{2}+|\partial A|$ where $t=t_{A}+t_{B}$.

We can emulate this effect in a global pure state by entangling the
triangles deep inside $A$ with those deep in $B$. Suppose we
begin with a particular product state $|\alpha\rangle$. We allow
ourselves to act on this state with single and two triangle unitary
transformations, say between any triangle and its neighbors.
On general grounds, a circuit of such one and two triangle unitaries
consisting of a polynomial in $t$ number of unitaries can produce
a state that looks effectively mixed for any $A$ with $t_{A}<t/2$.
This is simply the statement that thermalization can take place in
a time polynomial in the system size. However, note that this ``thermalization''
process has nothing to do with the physical dynamics, it is simply
a device to expose what sorts of superpositions in the ground state
subspace are reasonable.

We should also point out that states selected from the ground state
subspace with respect to the Haar measure on the projective space
also have this property of being effectively mixed for $t_{A}<t/2$ \cite{nature_ent}.
However, these states are not reachable from the product states $|\alpha\rangle$
with a polynomial in $t$ number of one and two triangle unitaries.
Hence we count them as unphysical since they could not be prepared
in a reasonable time for large $t$.

The entanglement properties of $A$ thus depend sensitively on $t_{A}$
and the choice of superposition in the ground state subspace. Note
that while we assumed that two triangle unitaries between neighboring
triangles should be allowed, in fact such neighboring triangles may
be quite distant from each other if the triangle density is going
to zero. Nevertheless, to obtain boundary law violations we must consider
superpositions obtained from such unitaries. Thus fix a graph $\Lambda_{3,t}$
with a given distribution of triangles and consider a family of regions
$A_{L}$ labeled by linear size $L$ such that $t_{L}\sim L^{a}$
with $d-1<a<d$. We assume that the total number of triangles and
sites is much larger than all region sizes considered. For a state
generated by a circuit of fixed depth consisting of one and two triangle
unitaries described above, there will be a range of region sizes $L_{\text{min}}<L<L_{\text{max}}$
for which the leading term in the entanglement entropy of $A_{L}$
is $L^{a}$. This entropy violates the boundary law and is due to
entanglement by construction (the global state is pure). $L_{\text{min}}$
is associated with lattice scale effects while $L_{\text{max}}$ reflects
the fact that a finite depth quantum circuit can only entangle triangles
a finite distance away. Once $L_{\text{max}}$ is reached, the entanglement
of $A_{L}$ due to correlations between triangles will be bounded
by $L^{d-1}L_{\text{max}}$ reflecting thermalization out to scale
$L_{\text{max}}$. The actual answer may be much less than this since
the number of triangles in $A$ is chosen to grow as a subextensive
power of $L$.

Although we focused on a particular realization of the triangle disorder, it is interesting to consider specifying a probability distribution for the disorder.  Suppose we introduce a probability measure $p(x_1,...,x_t)$ for the positions of the triangles in the lattice.  Let us assume this probability distribution is translation invariant so that $p(x_1+a,...,x_t+a) = p(x_1,...,x_t)$ for all $a$.  For concreteness, we imagine a large square lattice base graph with periodic boundary conditions in which triangles are placed.  If we further assume, as discussed above, that the states of interest look nearly maximally mixed for subsystems of size less than half the total system size, then the entanglement entropy of a subregion is simply proportional to the number of triangles contained within the subregion.  Thus to compute the average entanglement entropy of region $R$ we must average the counting function $C_R = \sum_i \chi_R(x_i)$ where $\chi_R(x)$ is one for $x \in R$ and zero otherwise.

The counting function has the property that $C_{AB} = C_A + C_B$, where $AB$ stands for the union of $A$ and $B$, provided $A$ and $B$ are disjoint.  Furthermore, the average $\bar{C}_R$ is clearly translation invariant $\bar{C}_{R+a} = \bar{C}_R$ because of the translation invariance of $p$.  These two properties are enough to prove that $\bar{C}_R \propto |R|$ i.e. the average entanglement entropy satisfies a volume law.  This is not terribly surprising given our assumption about the relevant superpositions in the ground state manifold.  For example, if the probability distribution $p$ is constant so that $p = (1/L^d)^t$ then we immediately find $\bar{C}_R = \frac{t}{L^d} |R|$. We thus conclude that to have any chance of obtaining something other than a volume law while maintaining translation invariance for the average entropy we would have to consider some other kind of superpositions in the ground state manifold. Note, furthermore, that our argument also suggests problems quite generally with having an anomalous scaling of the ``local ground state degeneracy'' with region size in a translation invariant system.

\subsection{Scoring the model}

We now briefly review our criteria and evaluate the successes and
failures of our model. We also answer a few other potential objections
to our construction.
\begin{itemize}
\item {{[}Short-range interactions{]}} Our Hamiltonian definitely consists
of only short-ranged interactions. However, we did have to consider
superpositions of ground states that can only be obtained from unitaries
coupling triangles that may be dilute.
\item {{[}Dimensionality{]}} When placed on a d-dimensional graph like
a hypercubic lattice, our model preserves the dimensionality of the
original graph. This is evidenced, for example, by the properties
of diffusion on the graph. It is true that a small number of links
are ``anomalous'', but there are not enough of these
links to alter the dimensionality of the diffusion problem.
\item {{[}Randomness{]}} Quenched randomness played a critical role in
our construction. The entanglement entropy depends sensitively on
which region we consider, but there are families of regions which
show anomalous entanglement entropy for a range of sizes. If we perform a disorder average to restore translational invariance, we recover a volume law for the entanglement entropy.
\item {{[}Fine-tuning{]}} Our model is expected to show fine tuning because
of the essential role of supersymmetry, however, from a field theoretical point of view we would not expect
to have to tune more than a finite number of supersymmetry breaking perturbations
to zero in an RG sense. Below we discuss the role of supersymmetry preserving and breaking perturbations on the level of the microscopic model in some detail. There is considerable ``tuning''
in the sense that we must select special subregions to see the anomalous
entanglement properties. It is not completely clear how to quantify
the extent to which our construction is generic.
\end{itemize}
Another possible objection to our model comes from the small and large staggering limits,
where it behaves somewhat like a self-contained system coupled to
$t$ free spins. However, we emphasize that this comparison is only
approximately valid, and in particular, at finite $y$ there are interactions
between the spins and the rest of the lattice. The power of supersymmetry
is that the exact ground state degeneracy remains even at $y$ of
order one where there is no notion of free spins. Thus the model we
construct is much less trivial than free spins sprinkled through an
otherwise conventional boundary law respecting system.

Indeed, more generally, the ground state degeneracy is robust against various supersymmetry preserving perturbations. First of all, the cohomology argument goes through unaltered if we modify the supercharges by including any non-zero site-dependent coefficients in their definition: $Q=\sum_i \lambda_i d_i^{\dagger}$ and $Q^{\dagger}$ its hermitian conjugate, where $\lambda_i$ are real and positive (phases can be absorbed in the definition of $d^{\dagger}_i$). So the ground state degeneracy is unchanged for this class of perturbations. Note that the staggering considered above, with $y>0$, falls into this class. It may be possible to construct other perturbations of the supercharges that leave the cohomology unchanged. Second of all, perturbations that leave the supersymmetry algebra and the Hilbert space unchanged will preserve the Witten index, as is clear from the definition:
\begin{equation}
W={\rm Tr} (-1)^F,
\end{equation}
where the trace runs over the entire Hilbert space. Its absolute value gives a lower bound to the number of ground states. For the decorated graphs with quenched disorder the Witten index is equal to the number of ground states, since all ground states have the same number of particles. It follows that supersymmetry preserving perturbations of this model that leave the Hilbert space unchanged can increase, but not reduce the ground state degeneracy. Due to the hard-core constraint, one can easily change the Hilbert space, for example by adding an edge, so in some sense the choice of the graph is fine tuning.

Finally, we should also comment on supersymmetry breaking perturbations.  In the spirit of our disordered model, we consider the possibility of adding supersymmetry breaking perturbations locally.  If we consider a dilute system of triangles, then local supersymmetry breaking perturbations should lift the degeneracy associated with triangles near such perturbations.  On the other hand, the large gap above the ground state manifold should not be immediately destroyed.  Indeed, in the limits of small and large $y$ and dilute number of triangles, the system is roughly equivalent to $t$ free spins in a non-uniform field i.e. each spin has its own magnetic field.  However, the case when the triangles are not dilute is potentially much more interesting since the ground state manifold can now support interesting dynamics below the gap of the unperturbed Hamiltonian.  It would be interesting to ask what sorts of linear combinations of the VBS states could be selected by such supersymmetry breaking perturbations and if the resulting combinations had any interesting entanglement properties.

\subsection{Comparison to holographic studies}
We briefly comment on the potential relations of our work to holographic studies.  First, the models we construct have degenerate ground states while holographically the situation is less clear.  The AdS$_2$ black hole certainly has an extensive ground state entropy, but this represents an extreme case in \cite{Dong12}.  Second, the ultimate low energy fate of the general class of models considered in \cite{Dong12} is often unknown.  As demonstrated by those authors, in some cases the IR fate of the RG flow is known and known to satisfy a boundary law.  More generally, it may be that the holographic computations should be interpreted as intermediate scale descriptions, a possibility discussed in \cite{Swingle11} where it was termed ``IR incompleteness''.  Certainly we cannot rule out ground states that display anomalous entanglement properties over a wide range of scales so long as they eventually satisfy a boundary law (up to logarithmic violations) at the largest scales.  This has been a central theme of our construction e.g. by focusing on the entangling power of finite depth quantum circuits and in the fractal Fermi surface example (see App. \ref{app:fractalFS}).

Another interesting issue is that it is unclear what state is being used in the holographic calculation.  In the case of the extremal black hole, the state is either the completely mixed state or a coherent superposition designed to mimic the mixed state for subsystems less than half the total system size.  This is analogous to the gravitational distinction between an eternal black hole and a black hole due to collapse in the finite past. The extremal black hole is a rather special object in this regard.  We showed above that both of these types of states can give anomalous entropy in our model, although it should only be called entanglement in the case of the coherent superposition.  The only type of state that does not give anomalous entropy in our model is a single localized ground state or non-generic superpositions of such states i.e. with triangles only entangled near the boundary.

One important distinction between our model and the holographic systems comes when considering the thermal entropy to entanglement entropy crossover.  We have an exponential number of strictly degenerate ground states separated from other states via a large gap. The temperature dependence of the thermal entropy will thus be of the usual type. However, the thermal physics in the holographic setup looks more like the codimension less than one scenario discussed in \cite{Swingle11} and, we suspect, like the thermal physics of the fractal Fermi surface over a wide range of temperatures. Since there are no known examples of field theories with codimension less than one, it seems that the ultimate IR fate of the ground state is to violate the boundary law at most logarithmically. In a sense this may also be the ultimate fate of our model if one assumes that a generic ground state must be connected to the product state via a quantum circuit of finite depth.

Finally, we note that the holographic models are manifestly translationally invariant. For the models we consider, however, it is easily seen that imposing translational invariance on the lattice leads to a ground state degeneracy that is exponential in the volume of the system. We also considered restoring translational invariance for the models with anomalous scaling of the ground state degeneracy by performing a disorder average and we found that this leads to a volume law for the entanglement entropy. 

\section{Conclusion}

The entanglement entropy of a many-body ground state of a local Hamiltonian typically obeys the area law. Violations of the area law may occur in the presence of gapless excitations, long-range interactions, infinite fine-tuning or ground state degeneracy. Gapless excitions, however, are believed to cause at most logarithmic violations of the area law \cite{Swingle11}. In this paper we presented a model with short-range interactions on a $d$ dimensional graph that can exhibit anomalous entanglement entropy scaling in a wide range of length scales in the absence of gapless excitations and infinite fine-tuning. The key ingredient to obtain this type of behavior is a controllable ground state degeneracy that results from quenched disorder and is protected by supersymmetry.

We showed that in the limits of small and large staggering there is a basis in which the degenerate ground states take on a simple form. In these limits the degeneracy can be understood in terms of local impurities and we found that these basis elements, which are maximally localized, each obey an area law for the entanglement entropy. We then showed that a generic ground state, which will typically be a superposition of the maximally localized states, can exhibit an anomalous entanglement entropy scaling. More precisely, the ground state degeneracy can be chosen such that the entanglement entropy of a generic ground state grows as $L^a$ with $d-1<a<d$ in a range of region sizes with linear size $L$. We carefully addressed the issues of what characterizes a generic ground states and what constraints the range of region sizes in which this anomalous entanglement entropy scaling is observed.

It is clear that this anomalous behavior occurs because the model we presented is not generic. We argue, however, that the ground state degeneracy arises in a reasonably sensible many-body Hamiltonian. We do this by scoring the model against a list of criteria that we impose on microscopic models of quantum many-body systems in order to be realistic and generic. The criteria concern the range of interactions, dimensionality, randomness and fine-tuning.

Finally, we discussed the possible relation of this work to holographic models that show anomalous entanglement entropy scaling of the same type. We believe an important question for the holographic models is for which state the holographic entropy computation is done. This is possibly related to the question what ultimately identifies the ground state of our model, or at least its entanglement entropy properties. Related to these issues is the question if there are systems for which anomalous violations of the area law persist on all length scales.

\begin{acknowledgments}
We thank Subir Sachdev for dicussions and, in particular, Shamit Kachru for sharing the results of \cite{Dong12} with us and for helpful discussions.  BGS is supported by a Simons Fellowship through Harvard University.  LH is supported by the Netherlands Organisation for Scientific Research (NWO).
\end{acknowledgments}

\bibliographystyle{unsrt}
\bibliography{bib2012,bibSUSY}

\appendix

\section{fractal Fermi surface}\label{app:fractalFS}
In this appendix we give a construction of a translation invariant fermionic state in any dimension that violates the boundary law for a range of region sizes.  For the violation to continue to the largest sizes we must consider long range hoppings.  The construction works in any dimension, but we consider two dimensions for illustrative purposes.  Thus consider a dispersion $\epsilon_k$ and chemical potential $\mu$ such that the Fermi sea $\Gamma$ has a Fermi surface $\partial \Gamma$ that is self similar in momentum space i.e. a fractal.  Of course, such a Fermi surface is highly non-generic, but given hoppings of range $\ell \gg 1$ on some regular lattice, we can tune the hoppings to produce features in momentum space on the scale of $1/\ell$.  Thus we expect to be able to approximate the fractal Fermi surface up to accuracy going like $1/\ell$ using hoppings out to a distance $\ell$.  Furthermore, if we take $\ell $ to infinity then it is plausible that we can arrange for a truly fractal Fermi surface, but we don't wish to contend with the increased mathematical subtlety of that situation here.

Suppose the Fermi surface curve may be approximately constructed in an iterative procedure i.e. like the Cantor set or the Sierpinski gasket.  After $n$ steps the length of the Fermi surface is $K_F(n) = \alpha^n K_F(0)$ while the feature size is $\Delta K = (1/\beta)^n K_F(0)$ with $\alpha,\beta >1$.  As we argued above, this pattern can continue at most until $\Delta K \ell \sim 1$ which bounds how large $n$ can be.  Thus for a range of scales in momentum space the Fermi surface looks self-similar, but for finite $\ell$ it eventually reduces to a conventional Fermi surface.  Thus we see elements both of long-range interactions and fine-tuning in this construction.

Now consider the entanglement entropy of a region $R$ of linear size $L$ in such a state.  For a conventional Fermi surface, the Widom formula for entanglement entropy, given by
\begin{equation}
S(L) = \frac{1}{12} \frac{1}{2\pi} \int_{\partial \Gamma} \int_{\partial R} |n_x \cdot n_k| \log{(L)}
\end{equation}
where $n_x$ and $n_k$ are unit normals to $\partial R$ and $\partial \Gamma$, is well established.  Let us now attempt to apply this formula in the case of the fractal Fermi surface.  First, if $L \gg \ell$, then the Fermi surface should look conventional and the Widom formula will predict the usual mild logarithmic violation of the boundary law.  Now consider the case $\ell \gg L$.  The degrees of freedom in a spatial region of linear size $L$ cannot distinguish wavelengths much longer than $1/L$ hence we expect that such a region can only resolve features in momentum space down to scales of order $1/L$.  Now a simple guess presents itself.  Let us use the Widom formula to estimate the entanglement entropy but using a Fermi surface which is smeared in momentum space on the scale of $1/L$.  In other words, we cutoff the interative construction when $\Delta K(n) L \sim 1$.

This cutoff defines $n(L)$ via
\begin{equation}
K_F(0) L = \beta^{n(L)}.
\end{equation}
Plugging into the formula for the length of the Fermi curve we obtain an effective length $K_F(L)$ as a function of region size $L$ given by
\begin{equation}
K_F(L) = K_F(0) \exp{\left(\frac{\log{\alpha}}{\log{\beta}} \log{(K_F(0) L)}\right)}.
\end{equation}
Defining $\gamma = \log{\alpha}/\log{\beta}$ we estimate the entropy as
\begin{equation}
S(L) \sim (K_F(0) L)^{1+\gamma} \log{L}.
\end{equation}
Thus for a range of scales $\ell > L > 1/K_F(0)$ we expect to find an anomalous violation of the boundary law for entanglement entropy.  This illustrates the power of long range hoppings and fine tuning for producing anomalous entanglement.  Note also that while this is not a rigorous argument, we believe it does correctly indicate a violation of the boundary law although we cannot vouch for the details of the scaling form.  This kind of fractal Fermi surface in some sense realizes the codimension less than one situation discussed in \cite{Swingle11} and effectively seen in holographic studies \cite{Dong12}

\end{document}